\documentclass[letter,scriptaddress,twocolumn,showkeys]{revtex4}

	\usepackage{amsmath}
	\usepackage[sort&compress]{natbib}
	\usepackage{makeidx}
	\usepackage{amsfonts}
	\usepackage[ansinew]{inputenc}
	\usepackage[usenames,dvipsnames]{pstricks}
	\usepackage{subfigure}
	\usepackage{epsfig}
	\usepackage{pst-grad} 
	\usepackage{pst-plot} 
	\usepackage[colorlinks,hyperindex]{hyperref}
	\hypersetup
	{
		colorlinks,%
		citecolor=blue,%
		linkcolor=blue,%
		urlcolor=black,%
	}

\newcommand{\beqn}{\begin{equation}}
\newcommand{\eeqn}{\end{equation}}
\newcommand{\beqna}{\begin{eqnarray}}
\newcommand{\eeqna}{\end{eqnarray}}
\newcommand{\nudyn}{$\nu_{\pm,\rm dyn}$}
\newcommand{\nudyncorr}{$\nu^{\rm corr}_{\pm,\rm dyn}$}
\newcommand{\Npart}{$N_{\rm part}$}
\newcommand{\elab}{$E_{\rm lab}$}
\makeindex
\begin{document}
\title{Net-charge fluctuation in Au+Au collisions at energies available at the Facility for Antiproton and Ion Research using the UrQMD model}
\author{Somnath Ghosh, Provash Mali, and Amitabha Mukhopadhyay}
\email{amitabha$_$62@rediffmail.com}
\affiliation{Department of Physics, University of North Bengal, Siliguri 734013, India}
\begin{abstract}
We have studied the dynamical fluctuation of net-charge of hadrons produced in Au+Au collisions at energies that in near future will be available at the Facility for Antiproton and Ion Research (FAIR). Data simulated by a microscopic transport model based on ultra-relativistic quantum molecular dynamics are analyzed for the purpose. The centrality and pseudorapidity dependence of the net-charge fluctuation of hadrons are examined. Our simulated results are compared with the results available for nucleus-nucleus collision experiments held at similar energies. The gross features of our simulated results on net-charge fluctuations are found to be consistent with the experiment. At incident beam energy $E_{\rm lab}=10A$ GeV the magnitude of net-charge fluctuation is very large, and in comparison with the rest its centrality dependence appears to be a little unusual. The effect of global charge conservation is expected to be very crucial at FAIR energies. The charge fluctuations measured with varying pesudorapidity window size depend on the collision centrality. The dependence is however exactly opposite in nature to that observed in the Pb+Pb collision at $\sqrt{s_{_{NN}}}=2.76$ TeV. 
\end{abstract}
\maketitle
\section{Introduction}
\label{intro}
Studies on event-by-event fluctuations of net-charge of the final state hadrons provide us with an opportunity to investigate the composition of the hot and dense matter prevailing the intermediate `fireball' created in high-energy nucleus-nucleus $(AB)$ collisions, which should in principle be characterized in the framework of quantum chromodynamics (QCD). It is argued that a phase transition from the quark-gluon plasma (QGP) to the usual hadronic state, considered to be an entropy conserving process, should produce such final states where the fluctuations of net electrical charge will be dramatically reduced in comparison with what is expected of an ordinary hadron gas (HG) system \cite{Jeon99, Jeon00, Asakawa00, Bleicher00, Heiselberg01}. The reason is, in a color deconfined extended QCD medium the charge $(q)$ carried by the elementary excitations are fractional $(q < |\mbox{charge of an electron}|)$ valued \cite{Ejiri06}. Event-by-event fluctuations of the net-charge or for that matter of any such conserved quantity, can also be directly related to the thermodynamic properties of the system created during the collision between two heavy nuclei at high energies. Hence they are expected to be affected by the location of the critical end point in a QCD phase diagram \cite{Stephanov98}. In that sense analysis of fluctuations of conserved quantities also helps us understand the structure of the QCD phase diagram.\\ 
\\
It is widely accepted that measurement of charge fluctuations is a useful tool to identify whether or not a deconfined QCD state is created in $AB$ collisions. All major collaborations on high-energy heavy-ion experiments have pursued such measurements on an event-by-event basis \cite{Adcox02-phenix,Adams03-star,Sako04-ceres, Alt04-na49, Abelev09-star, Abelev13-alice}. The observables for fluctuation measurements used by different groups may not always be the same, but the underlying correlation functions in all cases remain more or less identical. As introduced in \cite{Pruneau02}, experimentally the net-charge fluctuations are best studied by calculating the so-called dynamical net-charge fluctuation parameter (\nudyn). The quantity \nudyn\ is by definition free from the statistical effects, and most importantly it is independent of the detector efficiency. As a measure of the relative correlation strength between pairs of oppositely charged particles, \nudyn\, characterizes the nontrivial dynamics of particle production mechanism in $AB$ collisions. The formal definitions of \nudyn\, is given in section II of this article. In the STAR experiment \cite{Abelev09-star} held at the Relativistic Heavy Ion Collider (RHIC), it has been observed that at least in the region $19.6$ GeV $ \leq \sqrt{s_{NN}} \leq 200$ GeV, \nudyn\, is almost independent of the collision energy involved. The fluctuation measure however, is found to have a dependence on the  size of the colliding system. The CERES Au+Pb data on the other hand, showed that below $\sqrt{s_{NN}}= 17.2$ GeV the magnitude of \nudyn\, decreases with lowering energy \cite{Sako04-ceres}. So one may speculate that the dynamical net-charge fluctuation undergoes some kind of transformation as the intermediate fireballs produced in $AB$ collisions change from a baryon free to a baryon rich state. The STAR measurements are not consistent with the QGP hadronization model \cite{Jeon00, Heiselberg01}. Rather the estimated \nudyn\, values show a hadron gas like behavior in Au+Au and Cu+Cu collisions at $\sqrt{s_{NN}} \geq 19.6$ GeV. However, both the STAR and the CERES measurements qualitatively agree with the quark coalescence model \cite{Bialas02}, as well as with the resonance gas model \cite{Jeon00}. Recently the ALICE experiment held at the Large Hadron Collider (LHC) has reported their observations on \nudyn\, in Pb+Pb and p+p collisions at $\sqrt{s_{NN}} = 2.76$ TeV \cite{Abelev13-alice}. The ALICE p+p results follow the prediction of the hadron gas model, while the Pb+Pb results indicate that a QGP like state has perhaps been created.\\
\\
In this article we report a systematic study on the dynamical net-charge fluctuations in Au+Au collisions at energies expected to be available at the Facility for Antiproton and Ion Research (FAIR). In absence of any experimental data in this regard in this energy region, we have used the ultra-relativistic quantum molecular dynamics (UrQMD) code \cite{urqmd1} to generate particle emission data in Au+Au collisions at $E_{\rm lab} = 10A-40A$ GeV. In particular, we investigate the centrality and pseudorapidity interval size dependence of relevant fluctuation parameters. The collision centralities are determined by the Monte-Carlo Glauber model \cite{Miller07}. According to lattice QCD calculations \cite{Fodor03} at vanishing baryochemical potential $(\mu_b)$ the critical temperature $T_c$ required for a QGP-hadron phase transition should be around 170 MeV, and the corresponding critical energy density should be about a few GeV/fm$^3$. At FAIR the intermediate fireball is expected to be in the high $\mu_b$ side of the QCD phase diagram. Consequently, if a QGP-hadron phase transition takes place, the transition temperature would be significantly lower and the chemical potential significantly higher than the values required for a transition from a baryon free QGP state. Hopefully, the upcoming Compressed Baryonic Matter (CBM) experiment to be held at FAIR would be a witness to the not so well explored large $\mu_b$ region of the QCD phase diagram. Under these circumstances, simulation studies of known observables at and around the proposed FAIR energies, and comparison of these simulated results with whatever experimental results are available at comparable conditions are significant. Results of the present analysis may not be directly related to the QGP-hadron phase transition. Nevertheless, the study is certainly going to improve our understanding on various aspects related to charge fluctuation such as, the effects of a baryon rich environment on final state, contribution of conservation laws, extrapolation of observables to moderate collision energies etc. The present investigation is therefore expected to provide us with a basis to some important physics issues from the perspective of the CBM experiment \cite{CBM-Book}. 
\section{Dynamical net-charge fluctuations}
\label{Method}
The dynamical (non-statistical) fluctuations are generally determined by subtracting the statistical noise from the measured fluctuations. Assume that $N_{+}$ and $N_{-}$ are the multiplicities, respectively of the positively and negatively charged particles produced in an $AB$ event. The net charge and the total charge of the event are then denoted, respectively by $Q = N_+ - N_-$ and $N_{\rm ch} = N_+ + N_-$. Theoretically, the ratio of the variance of the net-charge $\delta Q^2$ and the total charge $N_{\rm ch}$
\begin{equation}
D = 4\, \frac{\left< \delta Q^2 \right>}{\left< N_{\rm ch} \right>}
\label{eq:D}
\end{equation}
is known as the $D$-measure. The $\left<~\right>$ symbol signifies that the quantity within these brackets is averaged over the ensemble of events considered. The variable $D$ takes care of the volume effect of the primary or secondary sources of particle production in $AB$ collisions. It also links the charge fluctuation with the entropy of the system. The measurement of $D$ was carried out using various models and under different circumstances e.g., with and without taking a QGP state into account, under a HG like situation \cite{Jeon00,Jeon99,Asakawa00,Jeon04,Pruneau02}, and considering various final state effects \cite{Zaranek02,Voloshin06}. For an uncorrelated gas of pions $D$ is estimated to be equal to $4$ \cite{Jeon99}. However, this value gets reduced if one considers the correlation between the positively and negatively charged particles produced from hadronization of gluons, quark-quark correlations, resonance decays etc. Considering particle correlations and resonance decays, lattice QCD calculation sets the limiting value as, $D \approx 3$ for a HG and $D \approx 1.0-1.5$ for a QGP like state \cite{Jeon04}. But the uncertainties in these estimations are always very large, and the major sources of them are the interactions between particles \cite{Abdel04, Shuryak01}, and the formulation of relating entropy to the number of charged particles in the final freeze-out state \cite{Jeon04}. In experiment $\nu_{\pm, \rm dyn}$, an experimentally accessible variable that is equivalent to $D$, is measured from the observation of the correlation strengths between identical and oppositely charged particle pairs. The variance of the difference between the relative number of positively and negatively charged particles is denoted by $\nu_{\pm}$ \cite{Pruneau02}, and it is given by
\begin{equation}
\nu_{\pm} = \left< \left( \frac{N_{+}}{\left< N_{+} \right>} - \frac{N_{-}}{\left< N_{-} \right>} \right)^2 \right>
\end{equation}
In the Poisson limit, i.e. for an independent emission of particles the above expression reduces to
\begin{equation}
\nu_{\pm,\rm stat} = \frac{1}{\left< N_{+}\right>} + \frac{1}{\left< N_{-} \right>}
\label{eq:NuStat}
\end{equation}
which is the statistical component of $\nu_{\pm}$.
The intrinsic or dynamical component of fluctuations is then obtained from the following formula,
\begin{equation}
\nu_{\pm,\rm dyn} = \nu_{\pm} - \nu_{\pm,\rm stat}
\end{equation}
A simple algebra leads to
\begin{equation}
\nu_{\pm,\rm dyn} = \frac{\left< N_+(N_+-1)\right>}{\left< N_+ \right>^2} +  \frac{\left< N_{-}(N_{-}-1)\right>}{\left< N_{-} \right>^2} -\frac{2\left< N_{-} N_{+} \right>}{\left< N_{+}\right>\left< N_{-}\right> }
\label{eq:NuDyn}
\end{equation}
The dynamical net-charge fluctuation is related to the $D$-measure through the following relation \cite{Jeon04,Pruneau02},
\begin{equation}
\left< N_{\rm ch} \right> \nu_{\pm,\rm dyn} = D-4
\end{equation}
The measurement of \nudyn\, is actually constrained by the conservation of global charges \cite{Pruneau02}, and in case of experiment also by the finite acceptance of the detector system used. If all the charged particles emerging from an interaction are measured, global charge conservation (GCC) would lead to vanishing fluctuation and the minimum value of the dynamical net charge fluctuation would then be $-4/\left< N_{4\pi} \right>$, where $\left< N_{4 \pi} \right>$ is the total number of charged particles produced over the $4\pi$ acceptance \cite{Pruneau02}. In this estimation it is assumed that charge conservation implies global correlations, but these correlations do not depend on the rapidity or pseudorapidity window size. After incorporating GCC into our scheme of things the corrected \nudyn\, reduces to
\begin{equation}
\nu^{\rm corr}_{\pm,\rm dyn} = \nu_{\pm,\rm dyn} + \frac{4}{\left< N_{4\pi} \right> }
\label{eq:NuDynCorr}
\end{equation}
while the corrected $D$-measure turns out to be
\begin{equation}
D^{\rm corr} = \left< N_{\rm ch} \right> \nu^{\rm corr}_{\pm,\rm dyn} + 4
\label{eq:DCorr}
\end{equation}
\section{The UrQMD model}
\label{model}
The model based on ultra-relativistic quantum molecular dynamics (UrQMD) \cite{urqmd1} is a microscopic transport theory that allows systematic studies of $AB$ collisions over a wide range of energies i.e., in the laboratory system from several GeV per nucleon up to the top RHIC energy ($\sqrt{s_{NN}} = 200$ GeV). UrQMD is based on the covariant propagation of constituent quarks and di-quarks. However, it has been accompanied by mesonic and baryonic degrees of freedom. At low energies, typically at $\sqrt{s_{NN}} \lesssim 5$ GeV UrQMD describes the interaction dynamics in terms of purely hadronic degrees of freedom, whereas at higher energies ($\sqrt{s_{NN}} \gtrsim 5$ GeV) excitation of color strings and their subsequent fragmentation into hadrons are taken into account. In a transport model an $AB$ collision is assumed to be a superposition of all possible binary nucleon-nucleon $(NN)$ collisions. An $NN$ collision is allowed if the impact parameter $b$ of collision satisfies the criterion $b < \sqrt{\sigma_{\rm tot}}/\pi$, where the total cross-section $\sigma_{\rm tot}$ depends on the isospin values of the interacting nucleons and on the collision energy evolved. The projectile and the target nuclei are described by a Fermi gas model, and therefore, the initial momentum of each nucleon is distributed at random between zero and the local Thomas-Fermi momentum. Each nucleon is described by a Gaussian-shaped density function and the wave packet of each nucleus is then taken as a direct product of single-nucleon Gaussian functions, i.e. the nuclear wave packets are not properly (anti)symmetrized. The interaction term in the model includes more than $50$ baryon and $45$ meson species. The QGP-hadron phase transition and/or any critical phenomenon is not explicitly incorporated into this model. However, in equilibrium the systematics followed by UrQMD can be interpreted by an effective Hagedorn type of equation of state \cite{Belkacem98}. Transport models like UrQMD can treat the intermediate fireball both in and out of a local thermal and/or chemical equilibrium. The model thus provides an ideal framework to simulate $AB$ collisions, where in the absence of any QCD driven critical phenomenon the hadronic degrees of freedoms are allowed to play a leading role. In this analysis we use the UrQMD (v3.4) code with default parameter settings in the laboratory frame of reference. 
\section{Results}
\begin{figure}
\centering
\includegraphics[width=0.5\textwidth]{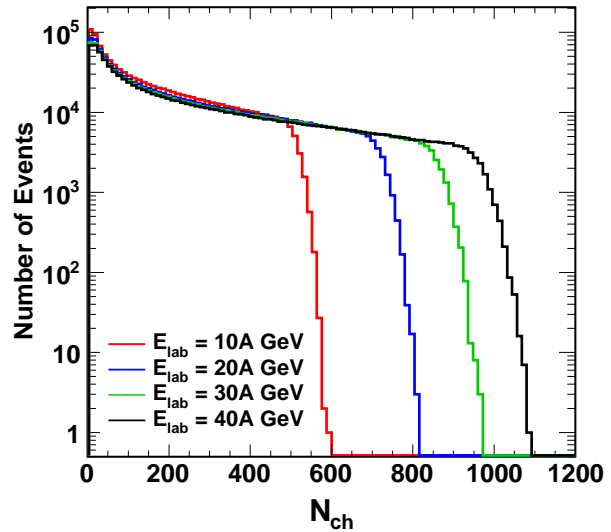}
\caption{(Color online) Charged particle multiplicity distributions in Au+Au collisions at $E_{\rm lab}=10A,~20A,~30A$, and $40A$ GeV.}
\label{MultiDist}
\end{figure}
\begin{figure}
\centering
\includegraphics[width=0.5\textwidth]{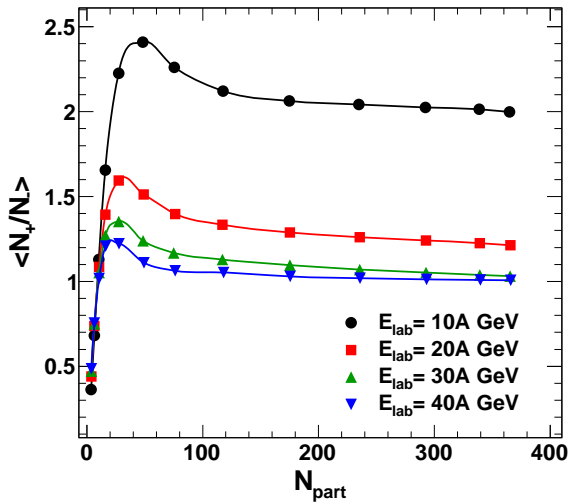}
\caption{(Color online) Ratio of positively and negatively charged particle numbers in Au+Au collisions at $E_{\rm lab}=10A,~20A,~30A$, and $40A$ GeV plotted against \Npart. The pseudorapidity width $\Delta\eta = 1.0$ about the centroid of the respective $\eta$ distribution, and $0.2 < p_t < 2.0$ GeV/c.}
\label{Ratio}
\end{figure}
\begin{figure}
\centering
\includegraphics[width=0.5\textwidth]{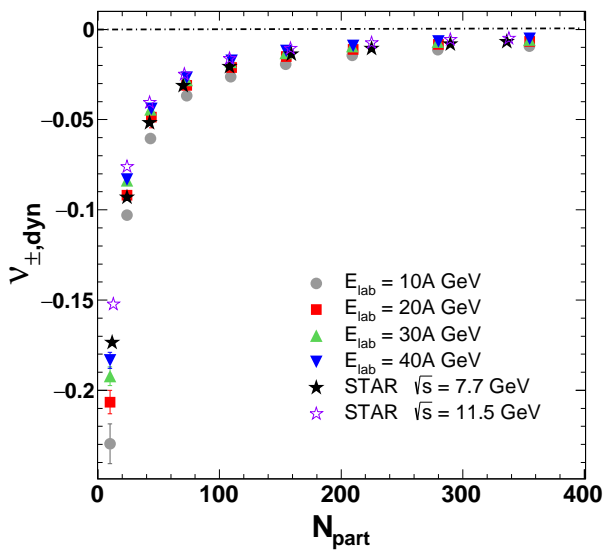}
\caption{(Color online) Dynamical net-charge fluctuations $\nu_{\pm,\rm dyn}$ is plotted against \Npart\, in Au+Au collisions at $E_{\rm Lab}=10A-40A$ GeV. The STAR Au+Au data at $\sqrt{s_{NN}} = 7.7 $ and $11.5$ GeV  are included from \cite{Sharma15}.}
\label{NuNpart}
\end{figure}
\begin{table*}
\caption{Average \Npart, average charged particle multiplicity $\left<N_{ch}\right>$ within $|\eta|<0.5$ and $0.2 \leq p_t \leq 2.0$ GeV/c, the dynamical net-charge fluctuations with and without GCC correction for some selective centrality bins in Au+Au collisions at $E_{\rm lab}=10A$ and $40A$ GeV. Note that the error values of \nudyncorr\, are almost identical to that of the corresponding \nudyn.}
\centering
\begin{tabular}{lllllll|lllll}
\hline\hline
& && $E_{\rm lab}=10A$   & GeV & &&    $E_{\rm lab}=40A$ & GeV\\
\cline{3-11}
Centrality(\%) &  $N_{\rm part}$ && $\left<N_{ch}\right>$  & ~~\nudyn && \nudyncorr & $\left<N_{ch}\right>$ & ~~\nudyn && \nudyncorr \\ [0.5ex]
 \hline
(0-5)\% & 343 && 143$\pm$0.065 & -0.0095$\pm$6.5$\times 10^{-6}$ && -0.0017 & 220$\pm$0.097 & -0.0052$\pm$0.6$\times10^{-6}$ && -0.0009 \\ 
(5-10)\% & 292 && 119$\pm$0.060 & -0.0111$\pm$7.2$\times10^{-6}$ && -0.0023 & 181$\pm$0.089 & -0.0061$\pm$1.2$\times10^{-6}$ && -0.0012 \\ 
(10-20)\% & 231 && 92$\pm$0.046 & -0.0134$\pm$0.1$\times10^{-6}$ && -0.0032 & 139$\pm$0.069 & -0.0080$\pm$3.4$\times10^{-6}$ &&-0.0020\\ 
(20-30)\% & 167 && 65$\pm$0.038 & -0.0180$\pm$0.2$\times10^{-6}$ && -0.0059 & 97$\pm$0.057 & -0.0108$\pm$3.2$\times10^{-6}$ &&-0.0031 \\ 
(30-40)\% & 118 && 45$\pm$0.032 & -0.0239$\pm$0.4$\times10^{-6}$ && -0.0097 & 66$\pm$0.047 &  -0.0158$\pm$1.0$\times10^{-5}$ && -0.0062 \\ 
(40-50)\% & 81 &&  30$\pm$0.026 & -0.0339$\pm$0.7$\times10^{-6}$ && -0.0174 & 44$\pm$0.038 & -0.0230$\pm$3.2$\times10^{-5}$ && -0.0110  \\ 
(50-60)\% & 52 &&  19$\pm$0.021 & -0.0492$\pm$1.8$\times10^{-5}$ && -0.0305 & 28$\pm$0.030 & -0.0356$\pm$6.9$\times10^{-5}$ && -0.0209 \\ 
(60-70)\% & 32 &&  12$\pm$0.017 & -0.0771$\pm$5.2$\times10^{-5}$ && -0.0564 & 17$\pm$0.023 & -0.0587$\pm$2.1$\times10^{-4}$ && -0.0412\\ 
(70-80)\% & 18 &&  ~\,6$\pm$0.012  & -0.1265$\pm$2.0$\times10^{-4}$ && -0.1040 & ~\,9$\pm$0.017 & -0.1020$\pm$8.2$\times10^{-4}$ && -0.0811\\ 
(80-90)\% & 9 &&   ~\,3$\pm$0.008  & -0.2433$\pm$1.2$\times10^{-2}$ && -0.2197 & ~\,5$\pm$0.011 & -0.1959$\pm$4.7$\times10^{-3}$ && -0.1736\\ 
\hline\hline
\end{tabular}
\label{table1}
\end{table*}  
As mentioned above the present analysis is based on Au+Au event samples generated by the UrQMD model at energies \elab = $10A$, $20A$, $30A$ and $40A$ GeV. Each sample consists of $10^{6}$ minimum bias events. The centrality of a collision is determined in terms of the total number of nucleons directly participating in the collision (\Npart), which is obtained from a Monte-Carlo Glauber model \cite{Miller07}. In Fig.\,\ref{MultiDist} we show the charged particle multiplicity $(N_{ch})$ distributions of the minimum bias event samples as mentioned above. The multiplicity distributions look more or less similar at all four \elab\, considered, excepting that the maximum value of $N_{ch}$ is higher at higher collision energy. $N_{ch}$ of an event plays an important role in the measure of charge fluctuation, as increasing multiplicity is expected to weaken the correlations among particles. Fig.\,\ref{Ratio} shows how the ratio of positively and negatively charged particle multiplicities varies with \Npart. Following the convention of the STAR experiment charged particles produced only within $\Delta\eta = 1.0$ about the central value of the respective $\eta$-distributions and within the transverse momentum $(p_t)$ interval $0.2 < p_t < 2.0$ are considered. One can see that in the central $\eta$-region the average multiplicity of positively charged particles is consistently larger than that of the negatively charged particles. At the beginning with increasing centrality the average ratio $\left<N_+/N_-\right>$ rapidly rises, reaches a peak, and then it slightly drops down only to attain a saturation in mid-central to central collisions. A lower collision energy results in a higher value of this ratio, the differences in saturation values however diminish with increasing energy. Excepting \elab\,$=40A$ GeV even the saturation values are greater than unity. The nature of $\left<N_+/N_-\right>$ against \Npart\, plot is guided by a dominance of baryons over antibaryons at this energy range. In Fig.\,\ref{NuNpart} the dynamical net-charge fluctuations \nudyn\, is plotted as a function of \Npart. The STAR Au+Au data at $\sqrt{s_{NN}} = 7.7$ and $11.5$ GeV are also included in this diagram. Once again we consider charged particles emitted within $\Delta\eta = 1.0$ about the peak value of the $\eta$-distribution and within $0.2 \leq p_t \leq 2.0$ GeV/c. We note that the \nudyn\, values are to be corrected for the limited bin effect i.e., the fluctuation measures are averaged according to the following averaging scheme,
\begin{equation}
\nu_{\pm,\rm dyn}(m_{\min} \leq m < m_{\max}) = \frac{\sum \nu_{\pm,\rm dyn}(m) P(m)}{\sum P(m)}
\end{equation}
Here $P(m)$ is the probability of having a total charge multiplicity $m$ and $\nu_{\pm,\rm dyn}(m)$ is calculated for the specific multiplicity $m$ using Eq.\,(\ref{eq:NuDyn}). With increasing centrality the observed trend of \nudyn\, for all the data sets seems to be almost identical -- sharply rising with increasing centrality in the extreme peripheral region, and subsequently attaining a saturation in mid-central and central collisions. Like experiments \cite{Adams03-star, Abelev09-star, Abelev13-alice} the correlation strength of the oppositely charged particle pairs i.e., the last term in Eq.\,(\ref{eq:NuDyn}) always dominates over the combined correlation strength of like charged particles, and hence at all centralities the dynamical fluctuations are negative. Though the energy dependence of \nudyn\, is marginal, for a given centrality bin it is seen that the magnitude $|\nu_{\pm,\rm dyn}|$ is a little higher at lower energy. For obvious reasons the magnitude of the dynamical net-charge fluctuations decreases if we take the GCC into account. In Table \ref{table1} we compare the values of \nudyn\, with the corresponding GCC corrected values (\nudyncorr) at $E_{\rm lab}=10A$ and $40A$ GeV and at some selective centrality intervals as a representative numerical output of our analysis. Note that the statistical errors in \nudyncorr\, are almost identical to that of \nudyn. It is seen that the charge conservation effect is more pronounced in the central collisions. After incorporating the GCC correction term the value of \nudyn\, increases by about 20\% in the 70-80\% centrality region, while the increment is about 80\% for the 0-5\% most central collisions. In the table we also quote the \Npart\, values along with $\left<N_{ch}\right>$ contained within $|\eta-\eta_0|<0.5$ and $0.2 \leq p_t \leq 2.0$ GeV/c for the chosen centrality bins. 
\begin{figure}[h]
\centering
\includegraphics[width=0.5\textwidth]{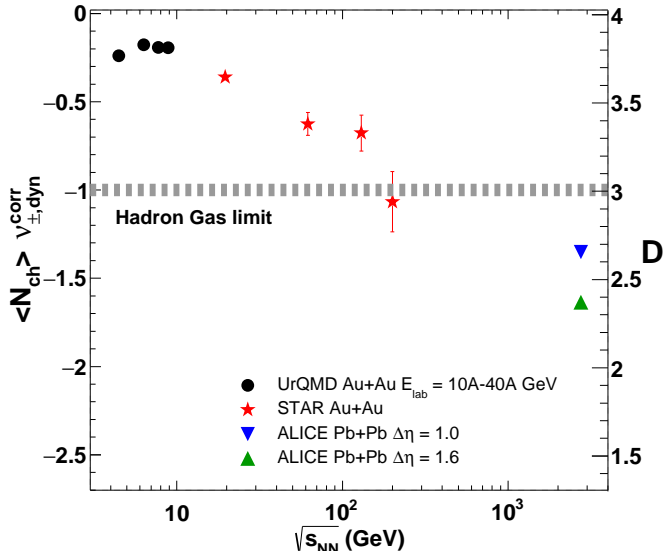}
\caption{(Color online) The energy dependence of $\left< N_{ch} \right>\,$\nudyncorr\,\,(left axis) and $D$ (right axis) for the (0-5)\% most central collisions. To match Eq.\,(\ref{eq:DCorr}) note that the variables are plotted with different scales. The STAR Au+Au and ALICE Pb+Pb results are taken from \cite{Abelev09-star, Abelev13-alice}.}
\label{NuEng}
\end{figure}    
The energy dependence of dynamical net-charge fluctuations is shown in Fig.\,\ref{NuEng}, where the STAR Au+Au result \cite{Abelev09-star} and ALICE Pb+Pb result \cite{Abelev13-alice} are included to depict the systematics followed by the fluctuation measure. In this figure we plot both $\left<N_{ch} \right>\,$\nudyncorr\,(left axis) and the $D$-measure (right axis) against $\sqrt{s_{NN}}$. The data points shown here are for the 0--5\% most central collisions and as before $\Delta \eta = 1.0$ centred about $\eta_0$. We see that barring an initial constancy, the fluctuation measure $\left< N_{ch} \right>\,$\nudyncorr\, decreases monotonically with increasing collision energy. Whereas the STAR Au+Au data at $\sqrt{s_{NN}}=200$ GeV is found to be very close to the prediction of a hadron gas, the ALICE result on Pb+Pb collisions at $\sqrt{s_{NN}}=2.76$ TeV and $\Delta\eta = 1.6$ shows significantly low fluctuations, indicating thereby that a probable source of dynamical net-charge fluctuation at LHC may be a QGP like state \cite{Abelev13-alice}. The STAR data points deviate upwards from the hadron gas limit with lowering in $\sqrt{s_{NN}}$. The UrQMD  prediction on $\left< N_{ch}\right>\,$\nudyncorr\, shows very little energy dependence within \elab\,$=10A-40$A GeV.
\subsection{Centrality dependence}
\begin{figure}
\centering
\includegraphics[width=0.5\textwidth]{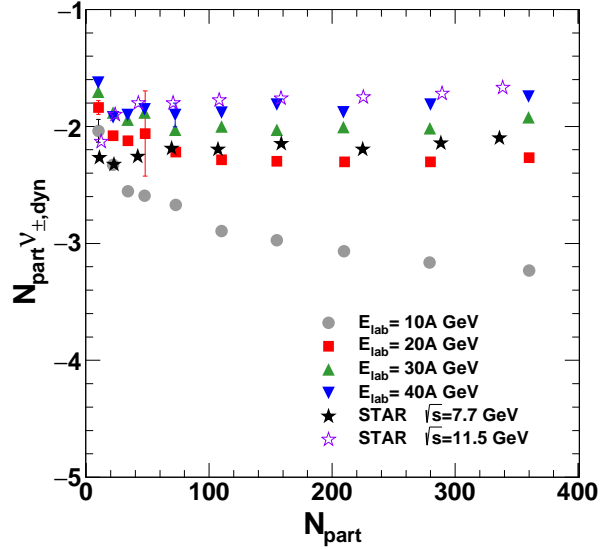}
\caption{(Color online) Centrality dependence of the dynamical net-charge fluctuations multiplied by the number of participating nucleons in Au+Au collisions. The STAR Au+Au results at $\sqrt{s_{NN}} = 7.7$ and 11.5 GeV are taken from \cite{Sharma15}.}
\label{NpartNu-Npart}
\end{figure}
\begin{figure}
\centering
\includegraphics[width=0.5\textwidth]{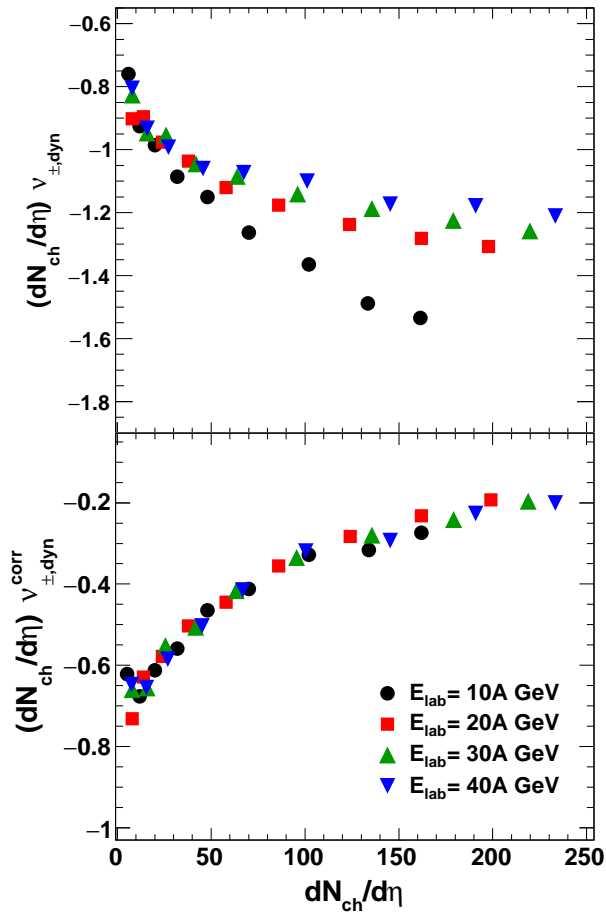}
\caption{(Color online) Dynamical net charge fluctuations multiplied by $dN_{ch}/d\eta$ are plotted against $dN_{ch}/d\eta$. The \nudyn\, and \nudyncorr\, values are considered in the top and bottom panels, respectively.}
\label{dNchNu-Nch}
\end{figure}
\begin{figure}
\centering
\includegraphics[width=0.5\textwidth]{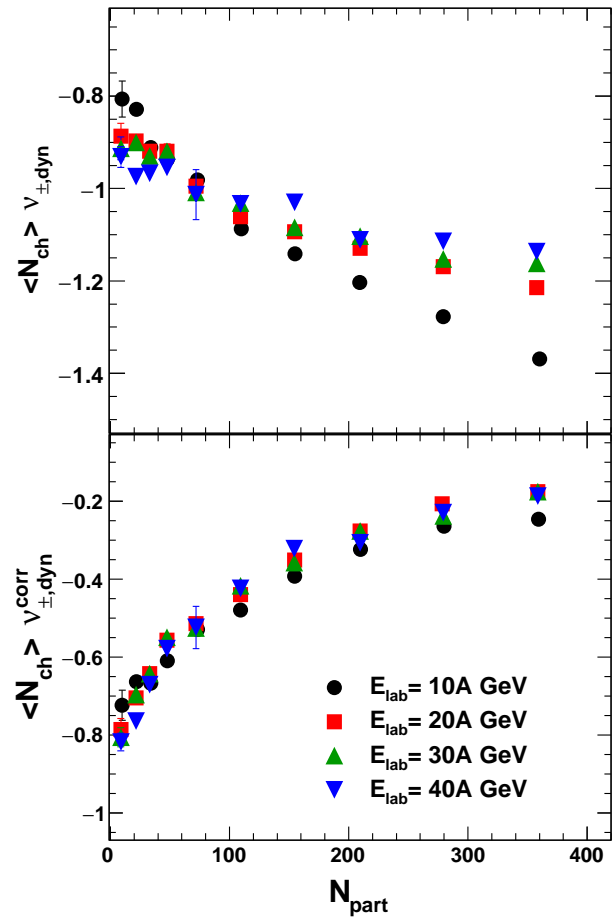}
\caption{(Color online) Centrality dependence of the dynamical net-charge fluctuations, multiplied by the average charge particle multiplicity $\left<N_{ch} \right>$. The product $\left< N_{ch} \right>\,\nu_{\pm,\rm dyn}$ (top) and $\left< N_{ch} \right>\,\nu_{\pm,\rm dyn}^{\rm corr}$ (bottom) are plotted against \Npart.}
\label{NchNu-Npart}
\end{figure}
The product of dynamical net-charge fluctuations and the number of participating nucleons is plotted against \Npart\,in Fig.\,\ref{NpartNu-Npart}, where the STAR Au+Au data \cite{Abelev09-star,Sharma15} at comparable energies are also included. Like the STAR results the UrQMD generated values of \Npart\,\nudyn\, at \elab\,$= 20A,\, 30A\,$ and $40A$ GeV are almost independent of the centrality of collisions, but the data show a definite energy ordering, i.e. for a given \Npart\, the values of \Npart\,\nudyn\, increase with collision energy. However, at $E_{\rm lab} = 10A$ GeV the UrQMD model simulated values of \Npart\,\nudyn\, monotonically decrease with increasing centrality of collisions. Such a deviation of the simulated data at $10A$ GeV has also been observed in some other plots shown in this article, particularly where the centrality dependence of \nudyn\, is involved. One probable reason for such deviation may be the low $\left<N_{ch}\right>$ value in $10A$ GeV events which increases the magnitude of \nudyn\, significantly. We know that at lower energies, typically $\sqrt{s}<5$ GeV ($E_{\rm lab} \lesssim 12.4A$ GeV), the UrQMD model describes the phenomenology of hadronic interactions in a way different from that at $\sqrt{s} \geq 5$ GeV. At energies $\sqrt{s}<5$ GeV the model takes the known hadrons and their resonances into account, and at $\sqrt{s} > 5$ GeV the excitation of color string and their subsequent fragmentation into hadrons are taken into account. As a result the multiplicity of charged hadrons produced in Au+Au collisions at $10A$ GeV collisions is considerably lower than the multiplicities at $E_{\rm lab} \geq 20A$ GeV. Moreover, the correction in \nudyn\, due to GCC is an important aspect, since the correction term is inversely proportional to $N_{ch}$.\\ 
\\
If an $AB$ collision is considered as a superposition of say $M$ nucleon-nucleon $(NN)$ sub-collisions, then one can write the one-particle density $\rho_1(\eta) = dN/d\eta$ as $\rho_1^{AB}(\eta) = M \rho_1^{NN}(\eta)$. In such a simplified scenario the invariant cross-section is proportional to the number of sub-collisions $M$, and the quantity \nudyn $dN/d\eta$ is independent of the centrality of collisions and also of the colliding system size. From the STAR measurements we have noticed that for Au+Au as well as for Cu+Cu collisions the magnitude of $|dN_{ch}/d\eta~\nu_{\pm, dyn}|$ increases by about 40\% as one moves from peripheral to central collisions \cite{Abelev09-star}. In Fig.\,\ref{dNchNu-Nch} we show a plot of the dynamical fluctuations multiplied by $dN_{ch}/d\eta$ as obtained from the UrQMD events at $E_{\rm lab}=10A-40A$ GeV as a function of charge particle density ($dN_{ch}/d\eta$) in the central particle producing region. The uncorrected and corrected values of \nudyn\, are plotted respectively, in the upper and lower panels of this diagram. We find that the magnitude of $dN_{ch}/d\eta\nu_{\pm,dyn}$ increases with increasing $dN_{ch}/d\eta$, and the change is measured to be approximately $40\%$ for the $20A-40A$ GeV collisions and $\sim 60\%$ for the $10A$ GeV collisions. Once again we see a large quantitative deviation in the $10A$ GeV result from the systematic behavior observed at other \elab. The energy dependence of the quantity $dN_{ch}/d\eta\,\nu_{\pm,dyn}$ within the energy domain of our analysis, is not very prominent. As mentioned, the GCC correction might be an important factor at lower energies where the produced charge particle multiplicity is not very high. This is depicted in the bottom panel of Fig.\,\ref{dNchNu-Nch}, where the values of $dN_{ch}/d\eta\,\nu^{\rm corr}_{\pm,dyn}$ are plotted against $dN_{ch}/d\eta$. The unusual deviation of  $10A$ GeV data points from the rest as seen in the upper panel of the figure has now disappeared, and the values of $dN_{ch}/d\eta\,\nu^{\rm corr}_{\pm,dyn}$ obtained from all four energies club together to increase monotonically with increasing $dN_{ch}/d\eta$. 
In Fig.\,\ref{NchNu-Npart} we plot the dynamical fluctuations multiplied by $\left<N_{ch}\right>$ contained within $\Delta \eta = 1.0$ against \Npart. As expected, the overall nature of the plots are identical to that shown in Fig.\,\ref{dNchNu-Nch}. Once again we observe a decreasing trend of $\left< N_{ch}\right>$\,\nudyn, while $\left< N_{ch}\right>$\,\nudyncorr\, increases monotonically with increasing \Npart. The deviation of the $10A$ GeV data points from the rest as seen in other diagrams, is removed when the GCC correction term is included in the expression of \nudyn.
\begin{figure}
\centering
\includegraphics[width=0.5\textwidth]{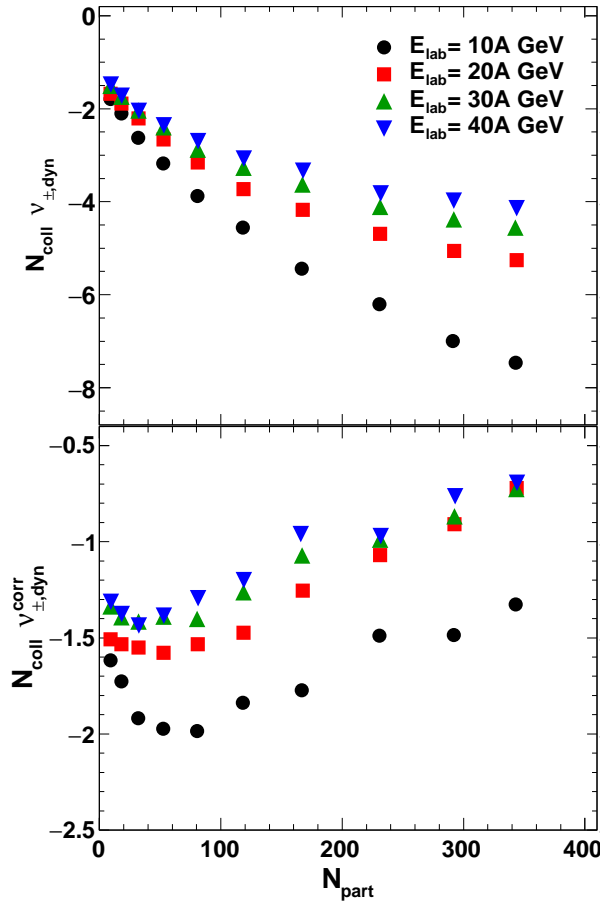}
\caption{(Color online) Centrality dependence of the dynamical net-charge fluctuations, multiplied by the number of binary collisions ($N_{\rm coll}$) in Au+Au collisions. The \nudyn\, and \nudyncorr\, values are considered in the top and bottom panels, respectively.}
\label{NcollNu-Npart}
\end{figure}
We now multiply the dynamical net-charge fluctuations by the number of binary collisions $N_{\rm coll}$ and plot the product against \Npart\, in Fig.\, \ref{NcollNu-Npart}. Once again we observe that the centrality dependence of $N_{\rm coll}$\,\nudyn\, is almost identical to the results shown in the upper panels of Fig.\,\ref{dNchNu-Nch} and Fig.\,\ref{NchNu-Npart}. Note that this type of centrality dependence of $N_{\rm coll}$\,\nudyn\, has also been observed in the STAR Au+Au data at $\sqrt{s_{NN}}= 62.4-200$ GeV and in Cu+Cu data at $\sqrt{s_{NN}} = 62.4$ GeV \cite{Abelev09-star}. On the other hand, with increasing centrality $N_{\rm coll}$\,\nudyncorr\, initially decreases up to $N_{\rm part} \sim 80$, and then increases monotonically. An energy ordering exists in this plot, and the values obtained from $10A$ GeV collisions are consistently and noticeably lower than those obtained from the other data sets. This is not very surprising, as particles produced in $10A$ GeV collisions are not of very high momentum, and these particles are not very much affected by binary collisions. Thus binary scaling of \nudyncorr\,is absent.
\subsection{Pseudorapidity interval dependence}
\begin{figure*}[t]
\centering
\includegraphics[width=\textwidth]{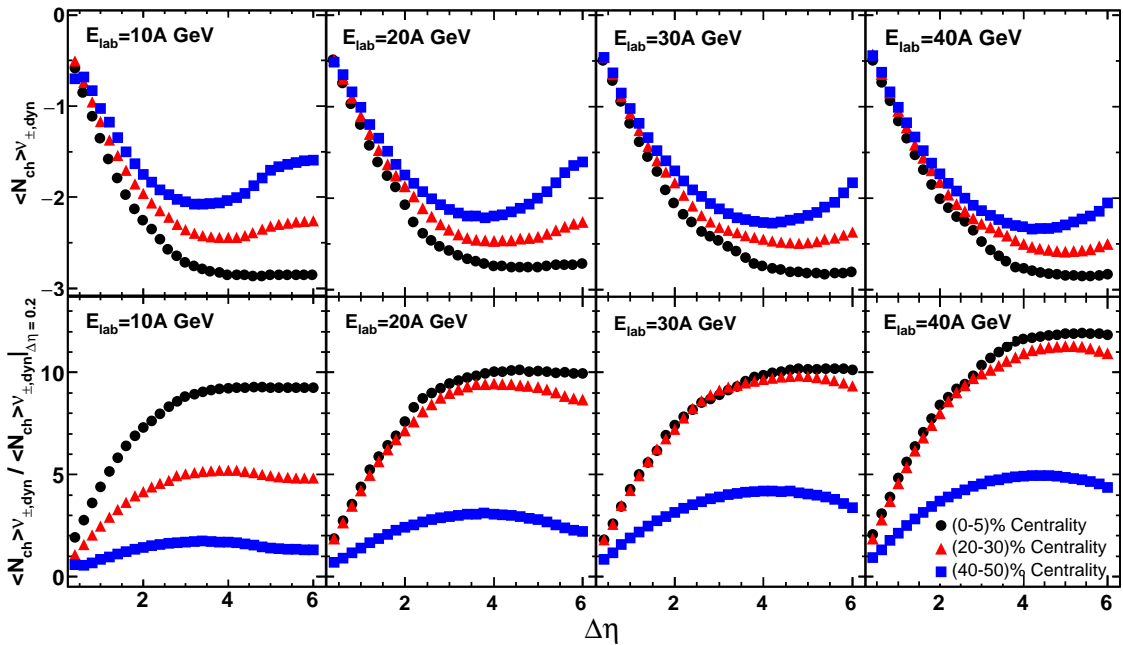}
\caption{(Color online) Values of $\left<N_{ch}\right>\nu_{\pm,\rm dyn}$ (top panels), and $\left<N_{ch}\right>\nu_{\pm,\rm dyn}$ relative to its value at $\Delta \eta = 0.2$ (bottom panels), plotted against $\Delta \eta$. The results shown are for centralities 0--5\%, 20--30\% and 40--50\%.}
\label{NuDelEta}
\end{figure*}
\begin{figure*}[tbh]
\centering
\includegraphics[width=\textwidth]{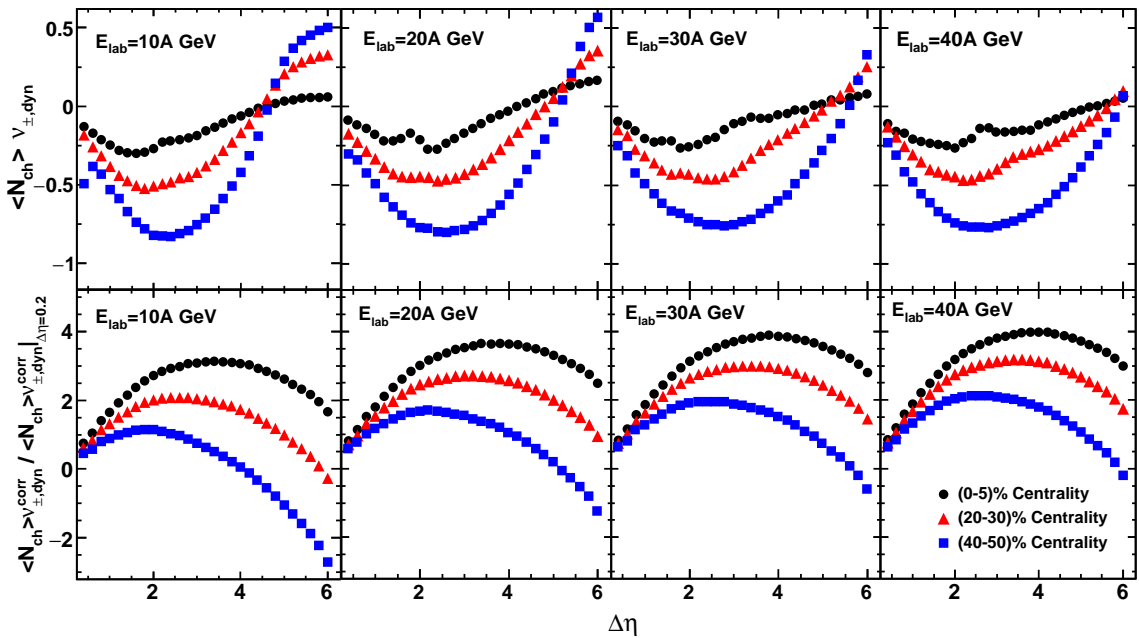}
\caption{(Color online) The same plots as shown in Fig.\,\ref{NuDelEta} but for the GCC corrected values of \nudyn.}
\label{NuCorrDelEta}
\end{figure*}
\begin{figure*}[t]
\centering
\includegraphics[width=\textwidth]{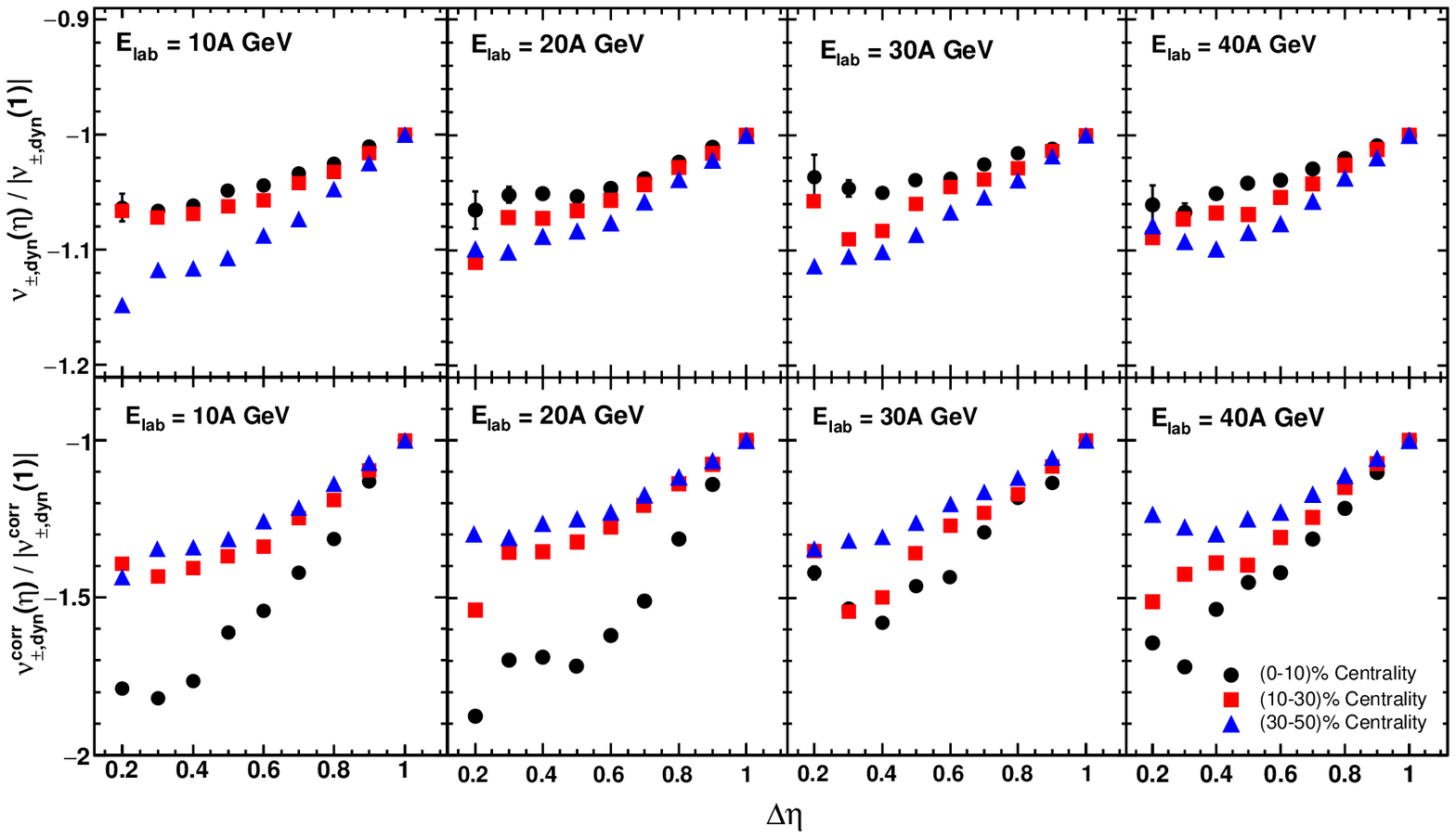}
\caption{(Color online) Values of $\nu_{\pm,\rm dyn}$ (top panels) and $\nu^{\rm corr}_{\pm,\rm dyn}$ (bottom panels) both relative to their magnitudes at pseurorapidity window $\Delta \eta = 1$, plotted against $\Delta \eta$. The results shown are for the 0--5\%, 20--30\% and 40--50\% most central collisions.}
\label{NuDelEta2}
\end{figure*}
The dependence of net-charge fluctuations on the pseudorapidity window size may provide us with some important information related to the hadronic medium created in high-energy $AB$ collisions on more than one counts. It helps to understand how the fluctuations evolve through a purely hadronic medium, and hence carves out the region of optimum coverage i.e. the pseudorapidity window size (say $\Delta \eta_{o}$) that is best suited for a fluctuation measurement. The net-charge fluctuations within a small $\Delta\eta$ may also be affected by global charge conservation. Charge fluctuations diffuse as the $\Delta\eta$ contains more number of particles, and at full $\Delta\eta$ coverage the net-charge fluctuations should vanish as the total charge of a system is conserved. On the other hand, the net-charge fluctuations measured within a narrow window may exclude some of the important aspects of the initial stage of the interaction. Thus an optimization of the window size is required for the measurement of charge fluctuations in high-energy heavy-ion collisions. In a recent simulation study based on the UrQMD generated Au+Au collision events at top RHIC energy, the optimum value of $\Delta \eta $ for \nudyn\, measurement is found to be $\Delta \eta_0 = 2.0-3.5$ \cite{Sharma15prc}. The $\Delta\eta$ dependence of GCC had been studied in \cite{Jeon00, Bleicher00} on the basis of the assumption that equilibration is established in the final stage of the collision. However, the suppression of net-charge fluctuations measured by the ALICE detector \cite{Abelev13-alice} indicates that in addition to GCC some other effect(s) such as the non-thermal effects \cite{Asakawa00}, are also important. The equilibration assumption has been revisited in \cite{Sakaida14} and it is found that fluctuations at an early stage of high-energy interaction can significantly alter the characteristics of the $(\Delta \eta)$ dependence of $\nu_{\pm,\rm dyn}$. Another point of concern is the radial flow of particles produced in heavy-ion collisions \cite{Jeon02,Voloshin06}. Radial flow is developed at an early stage of the collision, and it may be manifested in angular, transverse momentum, and longitudinal two-particle correlations in the final freeze-out stage \cite{Adams03}. As a matter of the fact, it is also important to investigate the impact of radial flow in two-particle correlation functions, especially in net-charge fluctuations and charge balance functions. In order to do that, it is however essential to have a comprehensive knowledge on how two-particle correlations scale with the pseudorapidity / rapidity window size.\\
\\
In the top panel of Fig.\,\ref{NuDelEta} we plot the product of the dynamical net-charge fluctuations and the average multiplicity of charged particles i.e., $\left< N_{ch}\right>\nu_{\pm,\rm dyn}$ as a function of $\Delta \eta$ for three centrality regions namely, 0--5\%, 20--30\% and 40--50\% most central collisions. In the bottom panels of the figure the same product relative to its value at $\Delta \eta = 0.2$ is plotted against $\Delta \eta$. It is seen that for all three centrality bins and within the energy domain of this analysis, the product $\left< N_{ch}\right>\,\nu_{\pm,\rm dyn}$ starting approximately at a value $-0.5$ keeps on decreasing monotonically up to $\Delta \eta \sim 4.0$. The rate of decrease depends on the centrality of collisions -- for the 0--5\% most centrality it is maximum and for the 40--50\% centrality it is minimum. Beyond $\Delta\eta \approx 4.0$ the values of $\left< N_{ch}\right>\,\nu_{\pm,\rm dyn}$ obtained from the 0--5\% most central collisions are found to be asymptotic in $\Delta\eta$, the 20--30\% centrality data show slight increasing trend, while the 40--50\% centrality data points show a faster increasing trend. Note that in the energy range $E_{\rm lab} =10A-40A$ GeV, the maximum coverage should be $\Delta\eta \sim 6.0$. The nature of the $\left< N_{ch}\right>\,\nu_{\pm,\rm dyn}$ against $\Delta\eta$ plots at smaller $\Delta\eta\;(<3.0)$, and their centrality dependences are almost identical to the results shown by the ALICE group for Pb+Pb collisions at $\sqrt{s_{NN}} = 2.76$ TeV \cite{Abelev13-alice}. The ALICE data were however GCC corrected. In addition, we observe a faint energy dependence in these diagrams. The mutual separation between the $\left< N_{ch}\right>\,\nu_{\pm,\rm dyn}$ versus $\Delta \eta$ plots corresponding to three different centralities considered, decreases with an increase in the collision energy. At full $\Delta \eta$ coverage the fluctuation measures are expected to vanish, which due to dominance of baryon number over antibaryons are not found in the energy region considered. However, when the $\left< N_{ch}\right>\nu_{\pm,\rm dyn}$ values are normalized by their values at $\Delta \eta = 0.2$, a completely different pattern is observed. The normalized values of $\left< N_{ch}\right>\,\nu_{\pm,\rm dyn}$ monotonically increase with increasing pseudorapisity window size up to $\Delta\eta \sim 4.0$, and then they remain constant at 0--5\% centrality, or continue to decrease at 20--30\% and 40--50\% centralities for the remaining parts of the diagrams. The centrality dependence of the normalized fluctuation measures tus gets inverted with respect to their unnormalized values. The separation between the plots corresponding to 0--5\% and 20--30\% centralities dramatically reduces in the $E_{\rm lab} = 20A-40A$ GeV collisions. The effect of GCC on $\left< N_{ch}\right>$\,\nudyn\, can be seen by comparing Fig.\,\ref{NuDelEta} with Fig.\,\ref{NuCorrDelEta}. In the latter case we consider the GCC corrected values of \nudyn. In the upper panels of Fig.\,\ref{NuCorrDelEta} one can see that the fluctuation measures $\left< N_{ch} \right>$\, \nudyncorr\, for all three centrality bins as well as for the selected energies, initially decrease with increasing $\Delta\eta$, but after a certain value of the window size (slightly different for the central and mid-central collisions) they revert back to zero. In some cases the values of $\left< N_{ch} \right>\,$\nudyncorr\, at large $\Delta\eta$ even shoot up above zero, indicating an anti-correlation between the positively and negatively charged particles. This might be due to the contribution of the spectator particles in \nudyn. A similar observation has also been made for UrQMD generated Au+Au collisions at $\sqrt{s_{NN}}=200$ GeV \cite{Sharma15prc}. We find that the centrality dependence of $\left< N_{ch} \right>$\, \nudyncorr\, is exactly opposite in nature to that of the ALICE Pb+Pb results at $\sqrt{s_{NN}} = 2.76$ TeV \cite{Abelev13-alice}. It is also to be noted that compared to the LHC and top RHIC energies at lower energies ($E_{\rm lab} = 10A$ GeV) the degree of nuclear stopping as well as the rescattering effects in central collisions are expected to be very high, which may adversely affect the two-particle correlation. Also the medium created in $AB$ collisions at LHC and top RHIC energies is almost baryon free \cite{Wong-Book}, whereas at lower energies a baryon dominated medium is expected which may modify the dynamics of net-charge fluctuations. The ratio between $\left< N_{ch}\right>$\,\nudyncorr and $\left< N_{ch}\right>$\, \nudyncorr$|_{\Delta \eta = 0.2}$ plotted against $\Delta\eta$ as shown in the lower panel of Fig.\,\ref{NuCorrDelEta} for all three centralities and for the four \elab\, values gives inverted parabola like curves. At large $\Delta\eta$ we see that the ratio, especially for the 40--50\% centrality, goes down to zero. This observation complements the results presented in \cite{Sharma15prc} for Au+Au collisions at $\sqrt{s_{NN}} = 200$ GeV. However, in \cite{Sharma15prc} the values of $\left< N_{ch} \right>$\,\nudyncorr$/\left< N_{ch} \right>$\,\nudyncorr$|_{\Delta \eta = 0.2}$ and/or $\left< N_{ch} \right>$ \,\nudyncorr\, were found to remain almost constant over a reasonable $\Delta\eta$ range called the optimum window size. In this analysis we do not find such an optimum $\Delta\eta$ value.\\
\\
The dependence of dynamical net-charge fluctuations on rapidity (pseudorapidity) window size may unveil the influence of radial flow in two-particle correlation \cite{Abelev09-star}. It is argued that by narrowing down the window size the charge balance functions in central $AB$ collisions relative to peripheral collisions, as claimed to be a sign of delayed hadronization due to the formation of a QGP like state \cite{Bass00}, might to some extent be a manifestation of radial flow developed in the intermediate fireball created in $AB$ collision. In general, such radial flow induces significant amount of angular, transverse momentum, and longitudinal two-particle correlations \cite{Bass00,Voloshin06}. A thorough study in this regard is however necessary to find out the exact contribution of radial flow to two-particle correlations. We here calculate \nudyn\, and \nudyncorr\, for several values of $\Delta \eta~(\leq 1.0)$ for three centrality bins namely 0--10\%, 10--30\%, and 30--50\%. For each centrality bin the values of \nudyn\, and \nudyncorr\, are then normalized by the magnitudes of the respective values measured at $\Delta \eta = 1.0$. The normalized values are then plotted against $\Delta \eta$ in Fig.\,\ref{NuDelEta2}. One can see that at small $\Delta \eta\,(\lesssim 0.3)$ the normalized dynamical fluctuations, especially the GCC corrected values, are slightly erratic in nature. However, with increasing window size the ratio monotonically rises to $-1.0$. Our observation in this regard is consistent with the STAR result reported in \cite{Abelev09-star}. As pointed out in \cite{Abelev09-star}, the slope of \nudyn\,(\nudyncorr) versus $\Delta \eta$ plot might be a proxy to the correlation length in pseudorapidity space. Larger slope corresponds to a shorter correlation length, which is associated with more central collisions and in larger systems. This observation has also been confirmed by the measurement of charge balance function \cite{Adams03}. In this analysis the GCC corrected values of the dynamical fluctuations are found to be consistent with the observations of the STAR experiment \cite{Abelev09-star,Adams03}. Thus one can speculate that a significant part of the two-particle correlation can perhaps be initiated by the radial flow in Au+Au collisions at FAIR energies. However, a systematic collision energy scan and system size dependence study are required before making any conclusive statement. 
\section{Conclusion}
A systematic study of some basic aspects of dynamical net-charge fluctuations of particles produced in Au+Au collisions at \elab$=10A - 40A$ GeV has been presented using data simulated by the UrQMD code. The centrality, collision energy and pseudorapidity window size dependences of various key parameters related to the net charge fluctuation are examined. In the context of upcoming CBM experiment such studies provide important clues about the fireball composition at FAIR energies. It is expected that the intermediate fireball produced in $AB$ collisions will be rich in baryons in the energy region considered in the present investigation. On the contrary the RHIC and LHC experiments have shown enough evidence of production of almost baryon free fireballs. However, in almost all cases we observe that the general trend of the results on net-charge fluctuations under these two widely differing temperature and density (or baryochemical potential) values are more or less similar and consistent with each other. Differences between the two on a few counts may be attributed to the huge difference in charge particle multiplicities in FAIR and in RHIC/LHC energies. In some cases only the \elab$=10A$ GeV results exhibit a little departure from the rest. This may be due to a dominance of number of baryons over other charged particles present in the colliding system. Even such departures are wiped out when the results are GCC corrected for complete pseudorapidity coverage. The present set of UrQMD results does not make it imperative upon us to assume presence of a QGP like state.


\begin{thebibliography}{99}
\bibliographystyle{plain}
\bibitem{Jeon99} S. Jeon and V. Koch, Phys. Rev. Lett. 83, 5435 (1999).
\bibitem{Jeon00} S. Jeon and V. Koch, Phys. Rev. Lett. 85, 2076 (2000).
\bibitem{Asakawa00} M. Asakawa, U. Heinz and B. M\"{u}eller, Phys. Rev. Lett. 85,
2072 (2000)
\bibitem{Bleicher00} M. Bleicher, S. Jeon and V. Koch, Phys. Rev. C 62, 061902 (2000).
\bibitem{Heiselberg01} H. Heiselberg and A. D. Jackson, Phys. Rev. C 63, 064904
(2001).
\bibitem{Ejiri06} S. Ejiri, F. Karsch and K. Redlich, Phys. Lett. B 633, 275 (2006). 

\bibitem{Stephanov98} A. M. Stephanov, K. Rajagopal and V. E. Shuryak, Phys. Rev. Lett. 81, 4816  (1998).
\bibitem{Adcox02-phenix} K. Adcox {\it et al.} (PHENIX Collaboration), Phys. Rev. Lett. 89, 082301 (2002).
\bibitem{Adams03-star} J. Adams {\it et al.} (STAR Collaboration), Phys. Rev. C 68, 044905 (2003).
\bibitem{Sako04-ceres} H. Sako {\it et al.} (CERES/NA45 Collaboration), J. Phys. G 30, S1371 (2004).
\bibitem{Alt04-na49} C. Alt {\it et al.} (NA49 Collaboration), Phys. Rev. C 70, 064903 (2004).

\bibitem{Abelev09-star} B. I. Abelev {\it et al.} (STAR Collaboration), Phys. Rev. C 79, 024906 (2009).
\bibitem{Abelev13-alice}  B. I. Abelev {\it et al.} (ALICE Collaboration), Phys. Rev. Lett. 110, 152301 (2013).

\bibitem{Pruneau02} C. Pruneau, S. Gavin and S. Voloshin, Phys. Rev. C 66, 044904 (2002).
\bibitem{Bialas02} A. Bialas, Phys. Lett. B 532, 249 (2002).

\bibitem{urqmd1} S. A. Bass {\it et al.,} Prog. Part. Nucl. Phys. 41, 255 (1998); M. Bleicher {\it et al.}, J. Phys. G 25, 1859 (1999).

\bibitem{Miller07} M. L. Miller, K. Reygers, S. J. Sanders and P. Steinberg, Annu. Rev. Nucl. Part. Sci. 57, 205 (2007).
\bibitem{Fodor03} Z. Fodor, S. D. Katz and K. K. Szabo, Phys. Lett. B 568, 73
(2003).
 
\bibitem{CBM-Book} {\it The CBM Physics Book}, edited by B. Friman, C. H\"{o}hne, J. Knoll, S. Leupold, J. Randrup, R. Rapp, P. Senger, Springer (2011).  
\bibitem{Jeon04} S. Jeon and V. Koch, In {\it Quark-Gluon-Plasma 3}, edited by R. C. Hwa, and X. N. Wang, World Scientific, Singapore (2004), p.430; arXiv:hep-ph/0304012v1.

\bibitem{Zaranek02} J. Zaranek, Phys. Rev. C 66, 024905 (2002). 
\bibitem{Voloshin06} S. Voloshin, Phys. Lett. B 632, 490 (2006).

\bibitem{Shuryak01} E. V. Shuryak and M. A. Stephanov, Phys. Rev. C 63, 064903
(2001).
\bibitem{Abdel04} M. Abdel Aziz and S. Gavin, Phys. Rev. C 70, 034905 (2004).

\bibitem{Belkacem98} M. Belkacem {\it et al.}, Phys. Rev. C 58, 1727 (1998).

\bibitem{Sharma15} B. Sharma (for the STAR Collaboration), arXiv:nucl-ex/00145v1.

\bibitem{Sharma15prc} B. Sharma, M. M. Aggarwal, N. R. Sahoo and T. K. Nayak, Phys. Rev. C 91, 024909 (2015).
\bibitem{Sakaida14} M. Sakaida, M. Asakawa and M. Kitazawa, Phys. Rev. C 90, 064911 (2014).

\bibitem{Jeon02} S. Jeon and S. Pratt, Phys. Rev. C 65, 044902 (2002).


\bibitem{Adams03} J. Adams {\it et al.} (STAR Collaboration), Phys. Rev. Lett. 90,
172301 (2003).

\bibitem{Wong-Book} C. -Y. Wong, {\it Introduction to High-Energy Heavy-Ion Collisions}, World Scientific (2004), p.265. 

\bibitem{Bass00} S. A. Bass, P. Danielewicz and S. Pratt, Phys. Rev. Lett. 85,
2689 (2000); S. Jeon and S. Pratt, Phys. Rev. C 65, 044902 (2002).



\end{thebibliography}
\end{document}